\documentclass{emulateapj}

\submitted{Accepted to ApJ 15 October 2013}
\shorttitle{Si isotope homogeneity of the solar nebula}
\shortauthors{Pringle et al.}

\begin{document}

\title{S\lowercase{i} isotope homogeneity of the solar nebula}

\author{Emily A. Pringle\altaffilmark{1,2}, Paul S. Savage\altaffilmark{1}, Matthew G. Jackson\altaffilmark{3}, Jean-Alix Barrat\altaffilmark{4}, and Fr\'{e}d\'{e}ric Moynier\altaffilmark{1,2}}

\altaffiltext{1}{Department of Earth and Planetary Sciences and McDonnell Center for the Space Sciences, Washington University in St. Louis, One Brookings Drive, St. Louis, MO 63130, USA; eapringle@wustl.edu, savage@levee.wustl.edu.}

\altaffiltext{2}{Institut de Physique du Globe de Paris, Universit\'{e} Paris Diderot, 75005 Paris, France; pringle@ipgp.fr, moynier@ipgp.fr}

\altaffiltext{3}{Department of Earth Science, University of California, Santa Barbara, CA 93109, USA; jackson@geol.ucsb.edu.}

\altaffiltext{4}{Universit\'{e} Europ\'{e}enne de Bretagne, Universit\'{e} de Brest, CNRS UMR 6538 (Domaines Oc\'{e}aniques), I.U.E.M., Place Nicolas Copernic, 29280 Plouzan\'{e} Cedex, France, Jean-Alix.Barrat@univ-brest.fr.}

\begin{abstract}
The presence or absence of variations in the mass-independent abundances of Si isotopes in bulk meteorites provides important clues concerning the evolution of the early solar system. No Si isotopic anomalies have been found within the level of analytical precision of 15 ppm in \textsuperscript{29}Si/\textsuperscript{28}Si across a wide range of inner solar system materials, including terrestrial basalts, chondrites, and achondrites. A possible exception is the angrites, which may exhibit small excesses of \textsuperscript{29}Si. However, the general absence of anomalies suggests that primitive meteorites and differentiated planetesimals formed in a reservoir that was isotopically homogenous with respect to Si. Furthermore, the lack of resolvable anomalies in the Calcium-Aluminum-rich Inclusion measured here suggests that any nucleosynthetic anomalies in Si isotopes were erased through mixing in the solar nebula prior to the formation of refractory solids. The homogeneity exhibited by Si isotopes may have implications for the distribution of Mg isotopes in the solar nebula. Based on supernova nucleosynthetic yield calculations, the expected magnitude of heavy-isotope overabundance is larger for Si than for Mg, suggesting that any potential Mg heterogeneity, if present, exists below the 15 ppm level. 
\end{abstract}


\section{Introduction}

Primitive meteorites (chondrites) are the remnants of the first planetary bodies to accrete in the solar system. Variations in the relative abundances of an element's isotopes in meteorites provide important information about the pre-solar chemical environment and the physical processes governing the formation and evolution of the solar system. The large majority of mass-independent isotopic variations (i.e. isotopic variations after correction for traditional thermodynamic isotopic fractionation, in which the amount of fractionation scales in proportion to the mass difference between isotopes) at the bulk-meteorite scale have been attributed to incomplete mixing of two (or more) isotopically distinct nucleosynthetic reservoirs within the solar nebula \citep{BIRCK04,WARREN11,FITOUSSI12,MOYNIER12}; however, some variations have also been attributed to chemical fractionation \citep{CLAYTON02,MOYNIER13}. In contrast, a lack of isotopic anomalies in bulk meteorites would suggest the complete homogenization of any distinct nucleosynthetic components, which resulted in a uniform early reservoir prior to planetesimal accretion. Such homogenization has previously been suggested by analysis of several elements at the bulk-rock scale, including Fe and Zn \citep{DAUPHAS08,MOYNIER09,WANG11,TANG12}.

Of the light elements, O, S, and Ca show mass-independent fractionation in bulk meteorites \citep{CLAYTON93,RAI05,RAI07,SIMON09}. However, the interpretation of O isotope anomalies is debated; variations may not reflect initial heterogeneity in the solar nebula since other mechanisms for mass-independent O isotope fractionation have been identified \citep{THIEMENS99,CLAYTON02,YURIMOTO04,LYONS05}. After O and Mg, Si is the third lightest element with the three stable isotopes needed to determine mass-independent effects, and Si is a major element in the terrestrial planets.

Silicon is composed of three stable isotopes: \textsuperscript{28}Si (92.23\%), \textsuperscript{29}Si (4.68\%), and \textsuperscript{30}Si (3.09\%). The most abundant isotope, \textsuperscript{28}Si, is a principle product of oxygen burning in massive stars, and is produced in core-collapse supernovae and, to a lesser extent, Type Ia supernovae \citep{TIMMES96}. The two heavier stable isotopes of Si, \textsuperscript{29}Si and \textsuperscript{30}Si, are secondary nucleosynthesis products and are mainly produced during carbon burning through the reaction of an $\alpha$ particle with \textsuperscript{25}Mg and \textsuperscript{26}Mg, respectively. Another production source of \textsuperscript{29}Si and \textsuperscript{30}Si is s-process neutron capture during He burning in asymptotic giant branch (AGB) stars. 

Large Si isotopic anomalies are found in various types of presolar grains \citep{ZINNER06,LEWIS13}, and in fact the presence of Si isotope abundances that are significantly different than solar Si ratios is one identifying characteristic of a presolar grain. These anomalies have been attributed to distinct nucleosynthetic sources of Si that condensed prior to the collapse of the solar nebula and therefore retained their individual isotopic signatures \citep{LODDERS05}. However, the extent to which such reservoirs were mixed during nebular collapse remains open for debate. The isotopic analysis of bulk meteorites can provide information about the level of homogenization of material within the solar nebula. Early searches for Si isotopic anomalies in bulk meteorites found only mass-dependent isotope fractionation \citep{MOLINI86}, but advances in Multi-Collector Inductively-Coupled-Plasma Mass-Spectrometery (MC-ICP-MS) have improved analytical precision by a factor of three. The current level of precision of 15 ppm (2$\sigma$) has made it possible to refine the search for inherited presolar nucleosynthetic heterogeneities that were preserved during accretion in the solar nebula.

This study presents high-precision Si isotopic measurements on a wide range of terrestrial and meteorite samples in order to quantify possible heterogeneities in Si isotope composition at the bulk-rock scale. At the current level of precision, Si isotopes display widespread homogeneity across bulk solar system materials, suggesting that inherited Si isotopic anomalies were erased through mixing in the solar nebula prior to planetary accretion. A possible exception is the angrites, which may exhibit small excesses of \textsuperscript{29}Si. We discuss the implications of these results on the distribution of Mg isotopes. 

\begin{figure}
\plotone{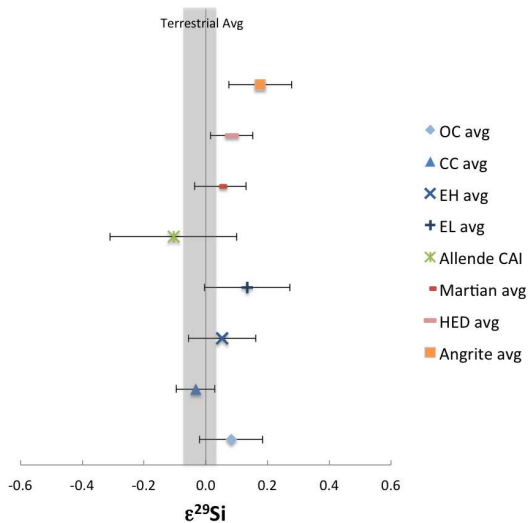}
\caption{Silicon isotopic compositions of terrestrial samples, bulk meteorites, and an Allende CAI. Data are given in $\varepsilon$\textsuperscript{29}Si (parts per 10,000) after internal normalization. Error bars represent the 2 standard error. The shaded box represents the average Si isotopic composition of all terrestrial samples measured ($\pm$ 2 se). Most planetary materials are within error of terrestrial composition, suggesting the homogeneous distribution of Si isotopes in the solar system.\label{fig1}}
\end{figure}

\section{Samples and Analytical Procedures}

This study reports the Si isotopic composition of a wide range of bulk solar system materials, including chondritic meteorites (5 ordinary chondrites, 8 carbonaceous chondrites, 6 enstatite H chondrites, and 7 enstatite L chondrites), achondritic meteorites (6 martian meteorites, 2 howardites, 9 eucrites, 3 diogenites, 5 angrites), and 16 terrestrial basalts. Achondrites represent the differentiated silicate fraction of planetary bodies. The achondrite samples reported here are believed to represent Mars, the asteroid 4-Vesta (in the case of the howardites, eucrites, and diogenites), and the angrite parent body. The terrestrial basalts include Ocean Island Basalts (OIBs) from Pitcairn and Samoa, which are locations representative of different mantle sources \citep[EM1 and EM2, respectively; ][]{HOFMANN97}. The Si isotopic composition of a Calcium-Aluminum-rich Inclusion (CAI) from the Allende carbonaceous chondrite is also reported.

The sample dissolution and chemical purification methods used are as described in previous Si isotope studies at Washington University in St. Louis \citep{PRINGLE13,SAVAGE13}. Powdered samples were dissolved in Ag crucibles using a NaOH alkali fusion technique and subsequently purified for Si isotope analysis through ion-exchange chromatography using BioRad AG50 X-12 (200-400 mesh) cation exchange resin, following the procedure developed by \citet{GEORG06}.

Silicon isotope compositions were measured on a Thermo Scientific Neptune Plus MC-ICP-MS at Washington University in St. Louis, operating in medium resolution mode. Measurements were made using standard-sample bracketing with the quartz sand standard NBS28 (NIST RM8546) subjected to the same Si purification procedure as the samples.

\section{Results}

\begin{deluxetable}{llrrr}
\tablewidth{0pt}
\tabletypesize{\small}
\tablecaption{Silicon Isotopic Compositions of Terrestrial Samples}
\tablehead{\colhead{Sample} & \colhead{Location} & \colhead{$\varepsilon$\textsuperscript{29}Si} & \colhead{2 se} & \colhead{n}}
\startdata
\textit{Terrestrial standard} \\
BHVO-2 &   & -0.02 & 0.06 & 115 \\
\textit{OIB} \\
OFU-04-14 & Samoa-Ofu & -0.11 & 0.12 & 10 \\
ALIA-115-03 & Samoa-Savai'i & -0.15 & 0.25 & 6 \\
ALIA-115-18 & Samoa-Savai'i & -0.08 & 0.12 & 6 \\
T16 & Samoa-Ta'u & -0.10 & 0.22 & 6 \\
T38 & Samoa-Ta'u & -0.06 & 0.10 & 6 \\
63-2 & Samoa-Vailulu'u & -0.21 & 0.08 & 6 \\
76-9 & Samoa-Malumalu & 0.14 & 0.14 & 6 \\
78-1 & Samoa-Malumalu & -0.06 & 0.19 & 6 \\
128-21 & Samoa-Taumatau & 0.12 & 0.19 & 6 \\
PIT 1 & Pitcairn & 0.04 & 0.09 & 6 \\
PIT 3 & Pitcairn & 0.03 & 0.16 & 6 \\
PIT 4A & Pitcairn & 0.04 & 0.19 & 6 \\
PIT 6 & Pitcairn & -0.12 & 0.23 & 6 \\
PIT 8 & Pitcairn & 0.08 & 0.12 & 6 \\
PIT 16 & Pitcairn & 0.02 & 0.21 & 6 \\
\textit{Average-Terrestrial} &   & -0.03 & 0.05 & 16 \\
\enddata
\end{deluxetable}

Silicon isotopic data are given in Tables 1-3 and Figure 1 in epsilon units (deviation in parts per 10,000 relative to the NBS28 standard; Equation (1)) after internal normalization to a \textsuperscript{30}Si/\textsuperscript{28}Si ratio of 0.03347 using an exponential law \citep{MARECHAL99},
\begin{equation}
\varepsilon^{29}\text{Si}=\biggl(\frac{(^{29}\text{Si}/^{28}\text{Si})_{\text{sample}}}{(^{29}\text{Si}/^{28}\text{Si})_{\text{NBS28}}}-1\biggr)10^{4}
\end{equation}
Errors cited are the 2 standard error (2 se; calculated as 2 standard deviation/$\surd$n) unless otherwise stated. 

Table 1 reports the Si isotopic compositions of the terrestrial basalts; both the replicate analyses of the basalt standard BHVO-2 ($\varepsilon$\textsuperscript{29}Si = -0.02 $\pm$ 0.06, 2 se, n = 115 measurements) and the combined average of all terrestrial samples analyzed ($\varepsilon$\textsuperscript{29}Si = -0.03 $\pm$ 0.05, 2 se, n = 16) are comparable to NBS28, suggesting that NBS28 is representative of a well-defined terrestrial Si isotopic composition. The group averages of all meteorite samples measured (with the exception of the angrites, discussed below) have Si isotopic compositions that are indistinguishable from the terrestrial average. 

The Si isotopic data for the chondrites and the Allende CAI are reported in Table 2. On average, the carbonaceous chondrites ($\varepsilon$\textsuperscript{29}Si = -0.03 $\pm$ 0.06, 2 se, n = 8), ordinary chondrites ($\varepsilon$\textsuperscript{29}Si = 0.08 $\pm$ 0.10, 2 se, n = 7), enstatite EH chondrites ($\varepsilon$\textsuperscript{29}Si = 0.05 $\pm$ 0.11, 2 se, n = 6), and enstatite EL chondrites ($\varepsilon$\textsuperscript{29}Si = 0.13 $\pm$ 0.14, 2 se, n = 7) all have Si isotopic compositions similar to terrestrial. The Si isotopic composition of the Allende CAI is also comparable to the terrestrial value within error ($\varepsilon$\textsuperscript{29}Si = -0.10 $\pm$ 0.20, 2 se, n = 9). 

\begin{deluxetable}{llrrr}
\tablewidth{0pt}
\tabletypesize{\small}
\tablecaption{Silicon Isotopic Compositions of Chondrites and Refractory Inclusions}
\tablehead{\colhead{Sample} & \colhead{Type} & \colhead{$\varepsilon$\textsuperscript{29}Si} & \colhead{2 se} & \colhead{n}}
\startdata
\textit{Ordinary chondrites} \\
Hallingeberg & L3.4 & -0.06 & 0.37 & 6 \\
Saratov & L4 & 0.12 & 0.19 & 7 \\
Tadjera & L5 & 0.20 & 0.20 & 7 \\
L'Aigle & L6 & 0.22 & 0.11 & 6 \\
Parnallee & LL3.6 & 0.01 & 0.18 & 13 \\
Olivenza & LL5 & -0.11 & 0.41 & 6 \\
Cherokee Springs & LL6 & 0.20 & 0.18 & 6 \\
\textit{Average-Ordinary chondrites} &   & 0.08 & 0.10 & 7 \\
\textit{Carbonaceous chondrites} \\
Orgueil & CI1 & 0.07 & 0.12 & 16 \\
Cold Bokkeveld & CM2 & -0.07 & 0.11 & 22 \\
Murchison & CM2 & -0.14 & 0.12 & 10 \\
Ornans \#1 & CO3.4 & -0.10 & 0.12 & 12 \\
Ornans \#2 & CO3.4 & 0.03 & 0.11 & 12 \\
\textit{Average-Ornans}&  & -0.04 &   & 2 \\
Lanc\'{e} & CO3.5 & 0.01 & 0.17 & 12 \\
Isna & CO3.8 & -0.17 & 0.18 & 10 \\
Allende & CV3 & -0.01 & 0.09 & 38 \\
Vigarano & CV3 & 0.07 & 0.14 & 11 \\
\textit{Average-Carbonacous chondrites} &   & -0.03 & 0.06 & 8 \\
\textit{EH chondrites}				 \\
Qingzhen & EH3 & 0.02 & 0.19 & 13 \\
GRO95517 & EH3 & 0.29 & 0.23 & 9 \\
Sahara 97076 & EH3 & -0.05 & 0.10 & 6 \\
Abee & EH4 & 0.03 & 0.13 & 14 \\
Indarch & EH4 & 0.10 & 0.29 & 13 \\
St. Marks & EH5 & -0.08 & 0.21 & 12 \\
\textit{Average-EH chondrites} &   & 0.05 & 0.11 & 6 \\
\textit{EL chondrites} \\
MAC 88184 & EL3 & -0.06 & 0.24 & 11 \\
Atlanta & EL6 & 0.08 & 0.22 & 12 \\
Hvittis & EL6 & 0.21 & 0.14 & 8 \\
Blithfield & EL6 & 0.47 & 0.32 & 3 \\
Eagle & EL6 & -0.06 & 0.12 & 9 \\
Khairpur & EL6 & 0.13 & 0.24 & 12 \\
LON 94100 & EL6 & 0.16 & 0.20 & 8 \\
\textit{Average-EL chondrites} &   & 0.13 & 0.14 & 7 \\
\textit{Refractory Inclusions} \\
Allende CAI &   & -0.10 & 0.20 & 9 \\
\enddata
\end{deluxetable}

Table 3 gives the Si isotope data for the achondrites measured in this study. The martian meteorite average ($\varepsilon$\textsuperscript{29}Si = 0.05 $\pm$ 0.08, 2 se, n = 6) is not resolvable from terrestrial Si isotopic composition at the current level of analytical precision. The Howardite-Eucrite-Diogenite (HED) meteorite group average falls within the terrestrial range but is very slightly offset from $\varepsilon$\textsuperscript{29}Si = 0 within 2 se ($\varepsilon$\textsuperscript{29}Si = 0.08 $\pm$ 0.07, n = 14); however, as no systematic variation is apparent, this is likely a result of an error underestimation. Finally, all the angrite samples have consistent \textsuperscript{29}Si excesses (on average, $\varepsilon$\textsuperscript{29}Si = 0.18 $\pm$ 0.10, 2 se, n = 5), suggesting a small non-linear isotopic effect. Taken together, all the sample groups (excluding angrites) give an average $\varepsilon$\textsuperscript{29}Si value of 0.06 $\pm$ 0.14 (2 standard deviation). We consider this error of $\sim$15 ppm (i.e., 0.15 epsilon units) in \textsuperscript{29}Si/\textsuperscript{28}Si after internal normalization as representative of the analytical precision for the present study.

\begin{deluxetable}{llrrr}
\tablewidth{0pt}
\tabletypesize{\small}
\tablecaption{Silicon Isotopic Compositions of Achondrites}
\tablehead{\colhead{Sample} & \colhead{Group} & \colhead{$\varepsilon$\textsuperscript{29}Si} & \colhead{2 se} & \colhead{n}}
\startdata
\textit{Martian meteorites} \\
Sayh al Uhaymir 008 & Shergottite & 0.14 & 0.22 & 7 \\
Los Angeles & Shergottite & -0.05 & 0.16 & 10 \\
Miller Range 03346 & Nakhlite & 0.18 & 0.17 & 6 \\
Lafayette & Nakhlite & 0.09 & 0.12 & 6 \\
Nakhla & Nakhlite & -0.03 & 0.20 & 6 \\
Allan Hills 84001 & Orthopyroxenite & -0.05 & 0.30 & 7 \\
\textit{Average-Martin meteorites} &   & 0.05 & 0.08 & 6 \\
\textit{HED} \\
Kapoeta & Howardite & -0.11 & 0.24 & 6 \\
Frankfort & Howardite & 0.18 & 0.18 & 5 \\
Serra de Mag\'{e} & Eucrite & -0.02 & 0.22 & 6 \\
Petersburg & Eucrite & -0.07 & 0.17 & 6 \\
Cachari \#1 & Eucrite & -0.04 & 0.44 & 6 \\
Cachari \#2 & Eucrite & 0.26 & 0.19 & 7 \\
\textit{Average-Cachari} &   & 0.11 &   & 2 \\
Camel Donga & Eucrite & -0.03 & 0.26 & 6 \\
Jonzac & Eucrite & 0.26 & 0.30 & 6 \\
Juvinas & Eucrite & 0.18 & 0.17 & 5 \\
Bouvante \#1 & Eucrite & 0.02 & 0.26 & 6 \\
Bouvante \#2 & Eucrite & 0.08 & 0.29 & 7 \\
\textit{Average-Bouvante} &   & 0.05 &   & 2 \\
Stannern \#1 & Eucrite & 0.00 & 0.26 & 6 \\
Stannern \#2 & Eucrite & 0.01 & 0.21 & 5 \\
\textit{Average-Stannern} &   & 0.01 &   & 2 \\
Pomozdino & Eucrite & 0.34 & 0.37 & 6 \\
Aioun el Atrouss & Diogenite & 0.04 & 0.21 & 6 \\
Tatahouine & Diogenite & 0.08 & 0.25 & 6 \\
Shalka & Diogenite & 0.15 & 0.12 & 6 \\
\textit{Average-HED meteorites} &   & 0.08 & 0.07	 & 14 \\
\textit{Angrites} \\
D'Orbigny &   & 0.27 & 0.14 & 12 \\
NWA1296 &   & 0.20 & 0.16 & 10 \\
NWA2999 &   & 0.05 & 0.13 & 10 \\
NWA4590 &   & 0.32 & 0.12 & 10 \\
NWA4931 &   & 0.13 & 0.15 & 10 \\
\textit{Average-Angrites} &   & 0.18 & 0.10 & 5 \\
\enddata
\end{deluxetable}

\section{Discussion}

The lack of any resolvable Si isotopic anomalies within the current level of precision at the bulk-rock scale in either primitive chondrites or meteorites derived from differentiated parent bodies (with the exception of angrites, see below) suggests that any inherited nucleosynthetic variations in Si isotopes were erased through mixing in the solar nebula prior to the formation of planetesimals. Therefore, the Si isotopic compositions of the different planetary bodies in the inner solar system represent a well-mixed average of different stellar sources. Furthermore, the single Allende CAI analyzed here has a Si isotopic composition that is indistinguishable from bulk chondrites and terrestrial samples, suggesting that the homogenization of Si isotopes occurred at the mineral scale as early as the age of the CAIs \citep[4.567-4.568 Ga;][]{BOUVIER07,JACOBSEN08,BOUVIER10}.

\begin{figure*}
\epsscale{0.75}
\plotone{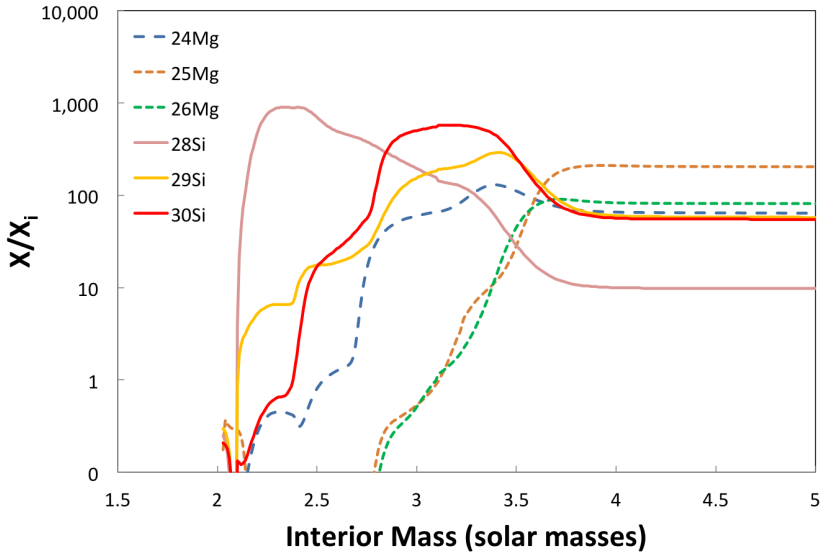}
\caption{Profile of Si and Mg isotope abundances (defined as mass fraction X relative to initial mass fraction X$_{i}$) as a function of interior mass coordinate in the supernova ejecta of the stellar model s25a37j of \citet{RAUSCHER02}.\label{fig2}}
\end{figure*}

The homogeneous distribution of Si isotopes has important implications for the distribution of Mg isotopes in the solar nebula. The presence of anomalies in Mg isotope abundances would have consequences for the \textsuperscript{26}Al-\textsuperscript{26}Mg short-lived isotope system (t$_{1/2}$ = 0.73 Myr), which is a principal chronometer for dating accretion and differentiation events in the early solar system \citep{JACOBSEN08,VILLENEUVE09,SCHILLER10}. A heterogeneous distribution of Mg isotopes in the pre-solar disk that was preserved in solar system materials during planetary accretion could lead to erroneous \textsuperscript{26}Al-\textsuperscript{26}Mg model ages. \citet{LARSEN11} found heterogeneities in the \textsuperscript{26}Mg excesses ($\delta$\textsuperscript{26}Mg\**) present in planetary samples with near-solar Al/Mg ratios. However, this heterogeneity can have two possible origins: (1) heterogeneous distribution of Mg isotopes or (2) variations in the \textsuperscript{26}Al/\textsuperscript{27}Al ratio within the solar nebula. The determination of Mg isotope anomalies attributable to incomplete nucleosynthetic source mixing is hindered by the input of \textsuperscript{26}Mg from \textsuperscript{26}Al radioactive decay, since the contribution of radiogenic \textsuperscript{26}Mg obscures the interpretation of nucleosynthetic isotope effects. 

The investigation of correlations in isotopic patterns in elements with related nucleosynthetic sources has been used to quantify possible heterogeneities in the early solar system \citep[e.g.][]{MOYNIER10}. Volatility differences may lead to a decoupling of elements with the same nucleosynthetic source. The chance of this thermal decoupling is minimized in elements with similar volatilities in the solar nebula, characterized by the 50\% condensation temperature (denoted T$_{c}$). Therefore, the comparable volatilities of Si and Mg \citep[T$_{c}$ Si = 1310 K; T$_{c}$ Mg = 1336 K;][]{LODDERS03} make Si isotopes well suited to constrain possible Mg isotope heterogeneity in the solar nebula.

Figure 2 shows Si and Mg isotope mass fractions as a function of interior mass in a 25 solar mass supernova ejecta of the stellar model s25a37j of \citet{RAUSCHER02}. Explosive carbon burning occurring in the outer layers of the supernova produces large excesses in \textsuperscript{29}Si and \textsuperscript{30}Si, while \textsuperscript{28}Si is mainly produced by oxygen burning in more interior regions. The pattern of Mg isotope production is different; \textsuperscript{24}Mg, \textsuperscript{25}Mg, and \textsuperscript{26}Mg are all co-produced during carbon burning. The outer regions of the supernova are only slightly enriched in the neutron-rich isotopes \textsuperscript{25}Mg and \textsuperscript{26}Mg, and, as a result, the magnitude of the heavy-isotope overabundance is smaller for Mg than for Si. Consequently, any excess in neutron-rich Mg isotopes would be associated with a much larger effect on neutron-rich Si isotopes. Following this logic, \citet{MOLINI84} suggested that the calculated Si isotopic effect would be 75 times larger than the Mg isotopic effect. The Si isotopic homogeneity of solar system material as indicated by this study (at $\pm$ 15 ppm) implies that Mg isotopes were also well mixed (at least at a 15 ppm level). Therefore, the heterogeneous distribution of \textsuperscript{26}Mg excesses across solar system materials \citep[e.g.][]{LARSEN11} must reflect heterogeneous distribution of the Al isotope ratio. 

The case of the small anomalies ($\varepsilon$\textsuperscript{29}Si = 0.18 $\pm$ 0.10, 2 se, n = 5) found in the angrites is puzzling. The mechanism by which a group of differentiated meteorites could preserve isotopic anomalies while all chondritic meteorites are isotopically homogenized is unclear. Use of the wrong power law for data normalization could introduce artificial isotopic effects in samples for which large mass dependent isotopic fractionation occurred due to a non-equilibrium (kinetic) process (e.g. evaporation). In this case, the fractionation would follow a Rayleigh distillation, and the generalized power law with exponent n = -$\onehalf$ would be the appropriate normalization scheme instead of the exponential case (generalized power law with exponent n = 0) used here \citep{MARECHAL99}. In other words, the Si isotopic fractionation between state a and state b can be expressed in terms of two fractionation factors $\alpha$ \citep{YOUNG04} such that
\begin{equation}
\alpha_{x/28}=\frac{(^{x}\text{Si}/^{28}\text{Si})_{\text{a}}}{(^{x}\text{Si}/^{28}\text{Si})_{\text{b}}}
\end{equation}
where x = 29 or 30, which are related by the expression
\begin{equation}
\alpha_{29/28}=\alpha_{30/28}\textsuperscript{$\beta$}
\end{equation}
where the exponent is expressed as
\begin{equation}
\beta=\frac{(1/m_{1}-1/m_{2})}{(1/m_{1}-1/m_{3})}
\end{equation}
in the case of equilibrium fractionation and 
\begin{equation}
\beta=\frac{\text{ln}(m_{1}/m_{2})}{\text{ln}(m_{1}/m_{3})}
\end{equation}
in the case of kinetic fractionation, and m$_{1}$, m$_{2}$, and m$_{3}$ are the masses of \textsuperscript{28}Si, \textsuperscript{29}Si, and \textsuperscript{30}Si, respectively. Numerically, the difference between equilibrium fractionation and kinetic fractionation changes the value of the exponent $\beta$ in Equation (3) from 0.5178 (equilibrium fractionation, Equation (4)) to 0.5092 (kinetic fractionation, Equation (5)).

The magnitude of the error introduced on $\varepsilon$\textsuperscript{29}Si through the use of an inappropriate normalization correction can be calculated from Equation (6), adapted from the example given for \textsuperscript{60}Ni by \citet{TANG12},
\begin{equation}
\varepsilon^{29}\text{Si}\approx5(n-k)\frac{\scriptstyle (29-28)(29-30)}{\scriptstyle 28}F\approx-0.09F
\end{equation}
where $n$ and $k$ are the power law exponents for Rayleigh or exponential isotopic fractionation, respectively (i.e., $n$ = -\onehalf, $k$ = 0), and $F$ is the isotopic fractionation in permil/amu. To account for the observed $\varepsilon$\textsuperscript{29}Si for the angrite average, a Si isotopic fractionation between 1 and 2 permil/amu (i.e., 10-20 $\varepsilon$-units on the \textsuperscript{29}Si/\textsuperscript{28}Si ratio) would be required. This exceeds by more than an order of magnitude the mass-dependent fractionation measured to date in bulk meteorite materials (which is currently less than 0.1 permil/amu, or 1 $\varepsilon$-unit on \textsuperscript{29}Si/\textsuperscript{28}Si). Therefore, the use of an inappropriate normalization scheme is an unlikely explanation to fully account for the Si isotopic anomalies in the angrite data. This suggests that angrites have preserved some small Si isotopic heterogeneity, the reasons for which are still unknown. These \textsuperscript{29}Si excesses may have implications for the distribution of Mg isotopes. Unfortunately, there is only one sample \citep[D'Orbigny; ][]{SCHILLER10} for which both Si isotopic data and the initial $\delta$\textsuperscript{26}Mg\** value are available, so it is impossible to interpret both isotopic systems at this stage. However, it has been noted that angrites have high initial $\delta$\textsuperscript{26}Mg\** compared to other differentiated meteorites \citep{SCHILLER10}, which could suggest some possible Mg isotope heterogeneity in line with \textsuperscript{29}Si excesses.  

\section{Conclusion}

After correction for mass-dependent isotopic fractionation, the Si isotopic composition of most meteorite groups is not resolvable from terrestrial composition within the current level of analytical precision. The lack of Si isotopic anomalies at the bulk-rock scale in solar system materials indicates homogenization of material in the solar nebula prior to planetary accretion. This constrains the possible heterogeneity of Mg isotopes to less than 15 ppm. 

\section*{Acknowledgements}

We thank Julien Foriel for maintaining the MC-ICP-MS and the clean lab facilities at Washington University in St. Louis and Francois-Regis Orthous-Daunay for assistance with the download of the stellar model data. This work was made possible by the provision of samples from Timothy McCoy (US National Museum of Natural History, Smithsonian Institution, Washington, DC), Alex Bevan (Western Australian Museum, Perth), Franz Brandstatter (Naturhistorisches Museum, Vienna), James Holstein, Clarita Nunez and Philip Heck (The Field Museum, Chicago), Meenakshi Wadhwa (Arizona State University, Tempe), Cecilia Satterwhite (NASA Johnson Space Center, Houston), the Comite de Gestion (Museum Nationale d'Histoire Naturelle, Paris), and Guy Consolmagno (Vatican). This work was supported by NASA EXO (NNX12AD88G) and COSMO (NNX12AH70G) grants and a chaire d'excellence from the Idex Sorbonne Paris Cit\'{e} to F.M. E.P. and P.S. thank the McDonnell Center for Space Sciences for funding.

\clearpage

\clearpage

\clearpage

\clearpage

\end{document}